\documentclass[fleqn]{article}
\usepackage{amssymb}
\usepackage[dvips]{graphicx}
\usepackage{amsmath}
\title{\Huge Three lectures on Newton's laws}
\author{Sergey S. Kokarev\thanks{logos-center@mail.ru}}
\date{RSEC "Logos"\,, Yaroslavl, Russia}

\begin{document}
\maketitle

\begin{abstract}
Three small lectures are devoted to three Newton's laws, lying in the foundation of classical mechanics. These laws are analyzed  from the viewpoint of
our contemporary knowledge about space, time and physical
interactions. The lectures were delivered for students
of YarGU in RSEC "Logos".
\end{abstract}

\section{Galileo's principle of inertia and\\ modern physics}

Coordinate method of describing physical phenomena is based upon two important and seemingly alternative notions.
The first is the idea that {\it all coordinate systems are equal in rights}, the second that {\it a coordinate system is being chosen
according to the practical convenience} arising when solving a specific physical problem.
When worded mathematically, that is in general covariant equations in terms of tensor bundles,
this seeming contradiction is resolved: invariant operations on tensors (tensor products, contractions, covariant differentiation)
do not depend on a coordinate system while their expression in components can be simplified if one chooses a coordinate system
which reflects existing symmetries.
Relativity theory allows us to place reference frames among four-dimensional coordinate systems where
reference frames world lines coincide with time coordinate \cite{vlad}.
Here again, as in the case of coordinate systems, ideas of {\it equality in rights of all reference frames} and {\it practical
convenience of a given system} can be applied. The correctness of the second can be easily established in
numerous examples from classic mechanics while the first idea is not so obvious in respect to Newton's mechanics.
Moreover, {\it Newton's first law}  also known as {\it the law of inertia}, states that there exist special reference frames
called {\it inertial frames} where mechanical bodies free from the influences of external forces persevere
in their state of being at rest or of moving uniformly straight forward.
As a rule, during a school or even college course of physics this law is seldom paid due attention to.
Sometimes, regrettably, one can find outright wrong interpretations of it.
Meanwhile, this law (that is better to call a principle, as we will show later),
when duly restated suddenly becomes a general principle valid not only for mechanics but for all modern physical-geometrical
theories of nature as well. For this reason we will examine its role in mechanics more closely.

Galilean law of inertia {\it postulates} existence of inertial frames of reference.
Our daily experience can tell us nothing about the motion of bodies not subject to external forces
or when these forces are balanced simply because we cannot isolate a body from the outside world.
Nor can we fully balance all the forces concerned (for logically there always remains a
possibility of  forces in existence perfectly unknown to us as yet) or exactly define {\it a geometrical straight
line} in  physical space. The latter aspect was pointed out by  H. Poincar\'{e} in 1909 \cite{poinc}.
It's evident that under such circumstances Galilean law of inertia {\it simply cannot be proven experimentally.}
That's the reason we stressed the law's postulative character in the beginning.

\begin{figure}
\includegraphics[width=0.6\textwidth]{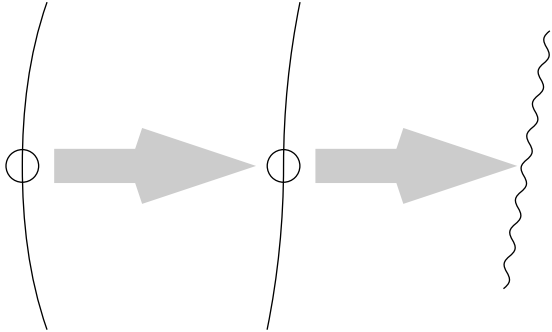}\caption{\small Changes in the path's scale of elaboration ---
both decreasing (to the left) and increasing (to the right) can show the path's deviations from a straight line.}\label{scale}
\end{figure}

But let us for the time being forget about the complication concerning the realization of
geometric straight lines. Let us assume that we have at our disposal
a certain device that can clearly and with any chosen degree of precision
show us whether the body in question moves straight forward or not,
without interfering with the motion of the bodies studied.
Upon eliminating the bodies "foreign"\, to our experiment and balancing the impacts of those remained,
if we find that the body in question moves straight forward {\it by any possible degree of
precision,}
then we can safely conclude that our reference frame is an inertial one, thus {\it having proved} Galilean
law of inertia experimentally. However, such outcome is hardly probable.
Experience shows that as the degree of precision increases there would {\it always}
appear new and more complex details of the experiment
which complicates the previous picture and ultimately leads to the new revisiting of the whole conception
behind it (Fig. \ref{scale}).

\begin{figure}
\includegraphics[width=0.4\textwidth]{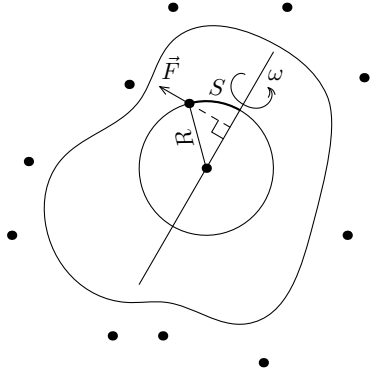}\caption{\small Hypothetical "fundamental force"\, of nature
on a planet with opaque atmosphere is in fact centrifugal force of inertia related to the planet's rotation.}\label{planet}
\end{figure}

Now we can almost ascertain that our device will show some deviations from a straight line.
But according to the law of inertia that means the body is either subject to some forces distorting
the path or our reference frame is not inertial. {\bf The choice between these two alternatives in not an easy one!}
Let us imagine a planet, Poincar\'{e} writes,
that rotates between relatively immobile stars and has an atmosphere opaque to the starlight (Fig. \ref{planet}).

Physicists of this planet are not able to observe their rotation around the stars and the resulting uninertiality
of their reference frame. Hence in their thorough experimentation they would consider the first alternative.
While studying the motion of different bodies they could at last arrive at a conclusion that deviations
from a straight line shown by test bodies indicate existence of a certain force that at two points on the surface ---
{}"repulsion poles"\, --- turns to zero, increases in proportion
$R\sin(s/R)$ where $s$ --- distancing from these points on the planet's surface, $R$ --- planet's radius;
this force is always directed at an angle $\pi/2-s/R$ to the vertical and is proportional to body mass.
The general formula looks like:
\begin{equation}\label{centrf}
|\vec F|=\alpha mR\sin(s/R)
\end{equation}
For the planet's inhabitants it would hold the same basic meaning as the law of gravity,
and $\alpha$ constant estimable by means of experiment would be considered one of nature's basic constants.
This state of affairs would last until it would dawn upon an alien Copernicus,
already familiar with the law of inertia as stated by an alien Galilei:
"May be there is no force $F,$ but the reference frame related to their planet is not inertial?
If we are to make the simplest assumption that the planet is in rotation
around an axe which pass through the "repulsion poles"\,, with constant angular velocity $\omega$, the expression for $F$
acquires a perfectly clear physical sense.
That would simply indicate the usual centrifugal force of inertia as daily observed by the planet's inhabitants,
only on a far more modest scale. In that case the formula (\ref{centrf}) should be rewritten as:
\begin{equation}\label{centri}
|\vec F|=m\omega^2r,
\end{equation}
where $r$ means distance from a point on surface to the rotation
axe\footnote{In the formula (\ref{centrf}) this distance is expressed by means of distance to the "repulsion poles"\,
on spherical surface. The point is, on the planet with constantly hidden stars,  inhabitants would have little use
of the system with parallels and meridians. Distances on the surface, on the contrary, would serve as a convenient coordinates.}.
And "fundamental constant of nature"\, $\alpha$ means squared angular velocity with which the surface rotates around some
invisible outward bodies with larger mass." Such or like would be the train of thought of our alien Copernicus.
For obvious reasons he would surely have to face serious trouble in persuading his fellow aliens to agree with this concept.

This example of Poincar\'{e}'s indicates the general way of reasoning characteristic for researchers who seek to test
Galilean law of inertia experimentally. Explaining of the deviations from a straight line would be done in terms of
interaction by means of forces; then, if it were possible, the problem would be reinterpreted in terms of uninertial
reference frames. Newton's first law allows both approaches. But which is better? Or, more precisely, which is more correct?
Before answering this questions let us first sum up. What do we gain by switching our focus from forces interactions to uninertiality?
Firstly, we eliminate the "excessive nature's forces"\,: in the example with the planet, after the discovery of our alien
Copernicus force $\vec F$ is "transferred"\, from dynamics to kinematics. Secondly, description of outside motions
in an open uninertial reference frame looks sufficiently simpler\footnote{Unfortunately, it is not a very sound argument for
the inhabitants of our twilight planet. For they have no picture of celestial bodies motion whatsoever!
The argument would get more sound with the appearance of  alien cosmonautics}!
Simplicity and beauty in this case are not mere abstract principles so dear to a philosophically inclined mind.
It is hardly coincidental that {\bf only after Copernicus' discovery Newton too could formulate a very simple and basic law of nature ---
namely, the law of gravity.} He was able to do that upon observing simple elliptic trajectories proposed by Kepler while
it would surely prove to be a task impossible for human mind  to arrive at the same conclusion through analyzing countless Ptolemaios' epicycles.
But it is beyond powers of today's computers, even most advanced among them, to perform inductive generalization, let alone grasp the notions of
beauty and simplicity!

Thus it should be apparent that the law of inertia is not a law in the proper sense of the word.
It says nothing of the world that Newton's classical mechanics described or attempted to describe.
It only {\it suggests a general rule of reasoning} which can be applied to this world.
In fact, {\it it wants us to assume that there are certain so called inertial reference frames, where the simplest situation
of free motion is represented by simplest possible kind of trajectory --- straight
line.}
It's a basic assumption of classical mechanics. Situation as described above {\it can be changed} only by forces or uninertial motions.
The decision which is the case can be made only after we've analyzed the whole situation, found forces at work,
estimated the possibility of some unknown forces interfering and defined whether the motion of outside forces looks simpler when described
in uninertial terms.

We can thus conclude that, its special status aside, {\bf the law of inertia is the most basic link of the whole logical structure of classical mechanics.}
In view of this we would like to stress that those authors who argue that Newton's first  law is the effect of the second are entirely wrong.
Their argumentation looks like follows. Let consider a particular case of the second law:
\begin{equation}\label{2Newton}
\ddot{\vec r}=\vec F/m,
\end{equation}
where $\vec F=0.$  Body moves with zero acceleration along a straight line, in full accordance with the law of inertia.
The mistake of this argument lies in the fact that Newton's second law in the form as cited above holds true only in inertial reference frames,
which existence is postulated in the first law. {\it Were it not for Newton's first law (and even were it otherwise formulated),
no one  would be able to write down Newton's  second law (or it would look absolutely different)!}
Were it not for  the first law, how could we know that the equation describing the motion of a free body looks like $\ddot{\vec r}=0,$
that it takes the form of a straight line equation!

\begin{figure}
\includegraphics{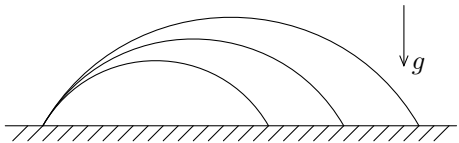}\caption{\small Trajectories of free particles in a world with modified Galilean
law of inertia. There is no gravitational force near the Earth's surface in this world.
As one moves further from the surface, gravity appears and increases and come close to a constant value
that equates $-m\vec g$ on great distances.}\label{parab}
\end{figure}

For example's sake let us imagine a slightly "modified"\, world where the law of inertia would read as
follows: {\it There exist special reference frames called inertial frames where mechanical bodies free from the influences of
external
forces move along parabolas with permanent vector coefficient in front of
squared
time.}
In latter case Newton's second law would look like that: $\ddot{\vec r}-\vec g=\vec F/m,$
where $\vec g$ stands for constant vector which sets the direction of the axes of all
parabolas.
Then, insofar as forces amount to zero we get $\ddot{\vec r}-\vec g=0,$
which means but parabolic motion (Fig. \ref{parab}).
If then we "recognize"\, gravitational acceleration in vector $\vec g$, then we'd  see that this "modified"\,
world is fairly identical with the world of Newton's classical mechanics near the earth's surface, but
all aspects of free fall and gravitation are being described without reference to forces.
What exactly did we do? By modifying Newton's mechanics, we've "transferred"\, one of the forces from dynamics to
free body kinematics, or, more precisely, to the forms of their trajectories.
Near the Earth's surface this kind of mechanics would be even more convenient than ours!
We could postulate some other curves instead of parabolas, and then some other forces would have left dynamics for free body kinematics.
When building his system of classical mechanics we've become so accustomed to,
Newton chose the most radical path: {\it made his free body kinematics simplest possible while his
force dynamics is as rich as it can be!} In this sense, Newton's form of mechanics is the simplest one can find.

This simplicity has the other end as well. For example, we could try to do it the other way round:
to convert the whole force dynamics into kinematics.  For gravity forces this problem has been already solved within
{\it general relativity theory} \cite{oto}, for other interactions there were more or less successful attempts at solving it
within {\it gauge principle of interactions} \cite{kalibr}. In both cases the curving of the paths is not caused by forces,
but comes from non-euclidian  geometry of the physical space-time which can have extra dimensions as well \cite{vlad1}.
Such "anti-newtonian"\, approach dubbed {\it the problem of geometrization of the physical interactions}\, \cite{grin}
grows still more popular within modern theories of physics.

As a conclusion we would like to show that there exist a quantum-mechanics analog of the Galilean
law of inertia. In nonrelativistic quantum mechanics we deal with states $|t\rangle,$
that are vectors of a certain Hilbert space and observables --- Hermitian operators acting on it \cite{dirac}.
General equation of nonrelativistic quantum mechanics:
\begin{equation}\label{schrod}
i\hbar\dot{|t\rangle}=\hat{\mathcal{H}}|t\rangle
\end{equation}
called the Schroedinger's equatuion is the quantum-mechanical analog of
Newtonian dynamics\footnote{This analogy takes more apparent form in a known Ehrenfest theorem for average values of classical dynamic qualities.}
equation (\ref{2Newton}).
In (\ref{schrod}) operator $\hat{\mathcal{H}}$
is the differential evolutionary operator, or Hamiltonian for short.
The transition to a common wave function $\psi(\vec r,t)$
is analogous to the choice of reference frames and transition from vectors to their projections in Newtonian classical mechanics.
Within quantum theory this procedure is called the transition to certain representation of a Hilbert space and operators in it.
Common wave functions are obtained via transition to the $\vec r$-representation: $\psi(\vec r,t)\equiv\langle\vec r|t\rangle.$
Now the quantum-mechanical "law of inertia'\, can be formulated as follows:

{\it "There are certain reference frames called inertial where the state of a free quantum particle of mass $m$ and momentum $\vec p$
in coordinate representation is described by a wave function proportional to $\exp[(\varepsilon t-\vec p\cdot \vec r)/\hbar],$
where $\vec p$ and $\varepsilon$ are connected by a standard relation: $\varepsilon={\vec p}\,{}^2/2m$"\,.}

It is this statement that allows us to concretize the Hamiltonian:
$\hat{\mathcal{H}}=\hat{\mathcal{T}}+\hat{\mathcal{U}}.$
With this in mind  we are able to prove that for the differential operators of the second order
the quantum mechanics law of inertia as formulated by us earlier, would lead to the only standard expression:
\[
\hat{\mathcal{T}}=-\frac{\hbar^2}{2m}\nabla^2,
\]
while the states described by plane waves thus appear
(in coordinate representation of the quantum mechanics) to be  quantum-mechanical analogs of the uniformly
straight forward motions of bodies in Newton's classical mechanics.

\newpage

\section{Two approaches to Newton's second law}

Newton's second law states that in inertial reference frames all
changes in the body velocity are caused by the influence of external forces\footnote{Here and below we assume that body mass remains
the same in the process of the motion. The case of bodies with changing mass can be always reduced to the motion of bodies
with constant mass, if we conceive that the mechanical system consists of subsystems the overall mass of which remains constant.}.
The aim of dynamics is to study the properties of nature's forces and to solve the problems concerning body motion under
influence of given forces. The vectorial equation of a point-like body motion has the form (\ref{2Newton}).
Let us consider a possibility of its being proven experimentally.
We've already discussed the conditional character that the notion of force has in the previous section.
Depending on the chosen formulation of the Newton's first law, some forces can "dissolve"\,
into kinematics and vise versa --- "crystallize"\, inside the dynamics.
Let us first accept the traditional Newtonian formulation of the Galilean law of inertia.
The experimental testing of the Newton's second law entails that during the experiment
we will probably need to measure {\it simultaneously and independently all three values $\vec a,\, \vec F,\,
m$}
of the expression (\ref{2Newton}) as well as substituting them into the expression (\ref{2Newton})
and to verify its identical fulfillment at the set level of precision.
We must also bear in mind that, as experience shows, in experiments of all kinds all our qualitative measurements are always
linked to the length measurement or the integer counting of time\footnote{Indeed, apart from measuring lengths with the help of a ruler
we use special rulers called clock-faces or scales of the multimeter, manometer, thermometer to measure time intervals, voltage,
pressure, temperature and so on. Measuring of time by the means of counting oscillations of a pendulum is an example of the
second type of quantitative measurements with help of whole counting. Pendulum's modern modifications ---
electric watch or counter of electric impulses device perform the same operation, only with the help of special electronic schemes.
Simple analysis shows that the work of all the other measuring devices is based on space scales or integer time counting.}.

\begin{figure}[bht]
\includegraphics{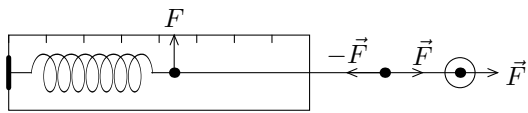}\caption{\small The work of the dynamometer is based upon the calibrating dependency (linear, as a rule)
and Newton's third law}\label{dinam}
\end{figure}

Thus even if we ignore the difficulties arising from the measuring of the derivatives as bo\-un\-da\-ry limit relations,
only acceleration remains to be measured by pure experiment. How then do we measure force and mass? Traditionally the force is measured
with the help of a dynamometer. But the work of a dynamometer (aside from the calibrating dependency specific for its type)
is based on the condition of the equilibrium of the bodies and Newton's third  law (Fig. \ref{dinam}).
If such a body remains in mechanical equilibrium while being connected to the dynamometer,
the force applied from the side of the dynamometer balances the measured force as applied by the outside bodies.
According to Newton's third law the same force is being applied by the body to the
dynamometer; when in equilibrium, its hand shows the force that is this outside force.
But to prove the third law in independent experiment we need some method for measuring forces.
Thus we come to a vicious circle, logically. In some textbooks authors use the experiment with
two weights on a thread that circulate around the center of mass, to illustrate the third law:
the relation of the weights centrifugal acceleration is inversely proportional to their mass relation,
therefore the product $ma$ would be the same by modulus for any two bodies interacting.
For our reasoning this kind of "proof"\, would not hold because it is based on the second law --- the same we're trying to prove.
The other way of "measuring"\, the forces would be via mass and acceleration. But then equation (\ref{2Newton})
becomes a mere definition of force. Besides, a simple analysis shows that any other methods of defining the mass will
inevitably rely on the second and third Newton's law (for example, rotation on a thread and weighing).

Then, if viewed logically, the "experimental"\, verification of the second Newton's law turns into a logical circle.
It doesn't mean much for practical purposes: Newtonian axioms are consistent and make it possible to pose and solve a lot of problems.
But forces and masses usually stay behind the scene, so to speak: all observations concern motions and trajectories.
Trajectories showing deviations from predicted on the basis of second law, which exceed the limits of possible error can
be explained either by the interference of some additional forces or existence of additional masses.

Scientists of different generations have pointed out this logical incompleteness in the experimental basis of the classical mechanics,
as well as the controversial status that force and mass have. Here are some examples:

\medskip
H. Herz: {\small "To find a logical fault in the system that has been worked out by the best of brains appears almost impossible.
But before we give up further investigation it is worth asking whether everybody,
the best of brains included, were satisfied by this system\dots As for me, I would especially point out the difficulty
one meets when explaining the introduction to mechanics to the thoughtful listeners, because here and there an excuse
is needed and one with some embarrassment hastens to turn to the examples which can speak for themselves". \cite{Herz} (English translation from
Russian text in \cite{kul}).}

\medskip
H. Poincar\'{e}: {\small "We meet the worst difficulties trying to define the basic notions. What is mass? ---
That, Newton says us, is the product of volume and density. ---
"Better said, that density is the quantity of mass in a unit of volume"\, --- answer Thompson and Tait.  ---
What is force? --- "That, Lagrange would say, is the cause that makes the body to produce motion or tend to produce motion".  ---
"That, answers Kirchhoff, is the product of mass to acceleration".
But why not say that mass is the quantity of force calculated for a unit of acceleration? This problem is unsolvable\dots
Thus we return to the definition given by Kirchhoff: force equates mass multiplied by acceleration.
This "Newton's law"\, ceases to be considered an experimental law and becomes a mere definition.
But the definition too is not sufficient for we don't know what is mass\dots
We've acquired nothing, all our efforts were in vain --- and thus we are forced to resort to the following definition which
only shows our defeat: masses are coefficients used in calculations for convenience.
\dots We have to conclude that it is impossible to give a fair idea what force and mass are within classical system"\, \cite{poinc2} (English translation from
Russian text in \cite{kul}).}

\medskip
A. Einstein: {\small "To connect force and acceleration becomes possible only after the new notion of mass is introduced ---
which, by the way, is explained through a seeming definition"\, \cite{ein}.}

It is clear that the source of our problems lies in the theory, not the experiments. Is it possible to formulate the laws of classical
 mechanics so as to escape the logical circle and to make the status of the quantities introduced clear? In effort to find the answer
 to this question we've arrived at two views on force and mass which will be discussed later.

1. {\bf Force and mass are superfluous notions and it is possible to formulate the laws of mechanics without
them.}

2. {\bf Force and mass are not two essences, but one which manifest itself
differently.}

\subsection{The operational formulation of the laws of mechanics}

First view should be attributed to the {\it operational approach} to the principles of physics \cite{oper}.
One of the operationalism's basic notions reads: {\it definitions of the physical quantities should be constructive,}
i.e. should in fact set the rules for measuring the quantity defined. Besides,
{\it the law of nature should be formulated in terms of the constructively defined physical quantities only.}
As we've discovered in previous section, the standard formulation of the classical mechanics is far from being operational.
Since acceleration is the only one constructively defined quality, the operational formulation of mechanics should be
based on the acceleration only. Below we give the operational formulation of the classical mechanics as stated by Y. I.
Kulakov,
who took it up within the scope of his
(meta)physical\footnote{The term "metaphysical"\, here is applied in a sense synonymic to the term "metamathematics"\,,
i.e. theory of cathegories, functors and toposes which allow to operate with  whole mathematical theories as mathematical objects of a higher order.
Therefore metaphysics is (or maybe will in future) a theory of physical theories.
Here we cite a fragment of the theory of physical structures by Y. I. Kulakov, slightly modified and adapted for our purposes.}
theory concerning physical structures \cite{kul}.
Within this formulation the basic notions are set $\mathfrak{B}$ of bodies and set $\mathfrak{F}$  of forces.
Let us examine the mapping $\mathfrak{B}\times\mathfrak{F}\to R,$ which image we will write down as $a_{i\alpha}$ for any $b_i\in\mathfrak{B}$
and
$f_\alpha\in\mathfrak{F}.$ We'll call $a_{i\alpha}$ acceleration\footnote{The vectorial aspect of this quantity here is irrelevant.
The reasoning below is true for one-dimensional motion. The three-dimensional case  is obtained via repetition of the reasoning
for the remaining two projections.}  of the body $b_i$ caused by the force $f_\alpha.$
Let us now consider an arbitrary body pair $\{b_i,b_j\}$ and force pair $\{f_\alpha,f_\beta\}.$
It appears possible to restate the Newton's second law in the terms of accelerations only:
\begin{equation}\label{2Newton1}
a_{i\alpha}a_{j\beta}-a_{i\beta}a_{j\alpha}=
\left|
\begin{array}{cc}
a_{i\alpha}&a_{i\beta}\\
a_{j\alpha}&a_{j\beta}
\end{array}
\right|=0.
\end{equation}
The expression (\ref{2Newton}) must be true for all bodies and forces from the sets $\mathfrak{B}$ and $\mathfrak{F}$ respectively.
We've formulated the second law in terms of accelerations, with the help of images of bodies and forces that have to satisfy the only condition --- to exist.
Qualitative characteristic of bodies (mass) and of forces (their values) are not included into the expression (\ref{2Newton1}).

Let us show now that the equation (\ref{2Newton1}) is in some sense equivalent to the equation (\ref{2Newton}).
In order to prove that we would need a mathematical theorem which was rigorously proven by G. G. Mikhailichenko in his series
of works devoted to the theory of physical structures \cite{mich}.
The theorem reads: {\it  if the equation (\ref{2Newton1})
is invariant in respect to the change of bodies and forces (the authors called this kind of invariance phenomenological symmetry),
then accelerations have the form:}
\begin{equation}\label{a}
a_{i\alpha}=\lambda_iF_\alpha,
\end{equation}
where $\lambda_i=\lambda(b_i),$ $F_\alpha=F(f_\alpha)$ are  some mappings $\mathfrak{B}\to R$ и $\mathfrak{F}\to R$ respectively.
It is easy to verify one side of the theorem by simple substitution of this form to the equation (\ref{2Newton1}).
The basic result of the theorem obtained via solving the complicated functional-differentional equations lies in the conclusion
(\ref{a}) from (\ref{2Newton1}). There remains only to recognize force $F_\alpha$ in the right side of the equation
(\ref{a}), as well as the reversed mass $\lambda_i=1/m_i$ sometimes called movability.
Therefore if the accelerations defined through sets of bodies and sets of forces satisfy the phenomenological invariant equation of the form (\ref{2Newton1}),
then bodies have masses, forces --- their values so that accelerations can be expressed through them by the relations of the form (\ref{a}).
{\bf This formulation of the Newton's law  follows as the rigorous mathematical concordance from the metalaw
(\ref{2Newton1}).}
The operational formulation of the classical mechanics was obtained, but at the "cost"\,
of a certain abstractization of the language and loss of visuality. It is interesting to note that for this formulation there's no need for
 the first Newton's law: set of forces includes forces of inertia, therefore the equation (\ref{a}) remains always correct.
Third Newton's law in its traditional formulation is not consistent with the operational formulation as described above:
forces and bodies belong to different sets, but the third law stated traditionally inevitably mixes these notions together.
However, with the help of an additional structure (algebra of bodies) it becomes possible to include the third law too:
it will be expressed by the additivity of the mapping $F$ in respect to the specific composition of bodies (see lecture 3).

\subsection{Classical mechanics as four-dimentional statics of the relativistic strings}

\begin{figure}
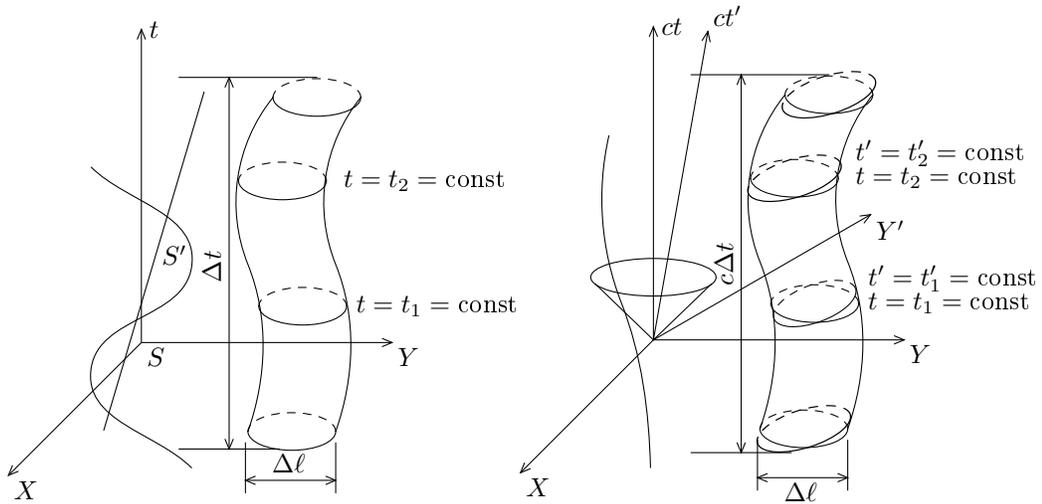

\includegraphics{allpic.5}\includegraphics{allpic.6}\caption{\small
Spaces of events in the classical Newtonian mechanics and SR. Diagram to the left shows the space line of a point-like body,
world line of some inertial reference frame $S'$, and a part of a world tube that represents a formal unity
of space positions of a certain three-dimensional body (shown by transversions $t=\text{const}$).
Diagram to the right shows the world line of a point-like body, axis $ct'$ and $Y'$ that moves in the direction
$OY$
of the inertial reference frame and a four-dimensional body --- relativistic world tube that has the same measure ---
four-dimensional length ---  in all directions. The sections of this four-dimensional body
into sequence of three-dimensional bodies depend on the choice of the reference frame. That's why it would be natural to
call world tube in SR {\it absolute history.}}\label{new}
\end{figure}

In order to explain the other approach to the laws of classical mechanics we would need to use some data
from the special relativity (SR) \cite{pauli}.
The SR's basic idea was first distinctly stated by Herman Minkowsky in 1909.
It postulates the affinity between space and time which, according to this theory, form the unite scene for
the events called {\it space-time.} Sometimes the SR's space-time is also called {\it four-dimensional Minkowsky space.}
In order to clear the difference between the absolute space and time of Newton's mechanics and the space-time of SR
let us look at the figures \ref{new}.

\medskip

\noindent
Both consist of elementary events ---
points with coordinates $(t,\vec r)$ in the first case and $(ct,\vec r)$
in the second. The motion of the point on both diagrams will be
represented by a certain curve (called world line): in the first case the curve's tilt to the time
axe can be at any angle (from null to $\pi/2$), in the second the tilt is limited to a cone of elements that tilt to the time axe at
angle $\pi/4$. The latter comes out of the finiteness of the maximal velocity that bodies and signals can acquire in SR,
i.e the finiteness of the velocity of light. Sections of the world lines by planes $t=\text{const}$
will result in a momentous position of a point in some reference frame fixed in space.
The motion of the extended bodies in both cases will be represented by world tubes, while the section of the world tubes by the lines $t=\text{const}$
would define the momentous position of these bodies in space (in a fixed space reference frame).
But the similarity between these two spaces of events remain at this purely visual level.
The time in classical mechanics is absolute and therefore, universal for all reference frames.
Figure to the left shows two reference frames: conditionally "stationary"\, $S$ and conditionally "moving"\, $S'$.
Intervals of time being counted by clocks in $S$ and $S'$ are identical regardless of their movement character and can be defined
on this diagram by projecting them on the axe of absolute time unique for all reference frames and bodies.
The diagram to the right shows the relativity of time in SR ---
there exist infinite set of times, every one of which moves in its own direction in four-dimensional space of time.
The intervals between a pair of events depend on the choice of a time line, i.e. reference frame and are related to each other through
specific SR formulae.

The difference between the diagrams becomes even more apparent if we look at the interval between events. In classical mechanics
it can be calculated only for simultaneous pairs of events while in space-time of SR distance (4-dimensional interval)
is defined for any pair of events. If in some reference frame the events got divided by a time interval $\Delta t$ and distance interval $\Delta l,$
the four-measure distance between these events can be defined by the formula:
\begin{equation}\label{int}
\Delta s^2=c^2\Delta t^2-\Delta l^2.
\end{equation}
The value of the interval does not depend on the choice of the reference frame. That allows us to deduce the laws of transition from one
 reference frame to another (Lorenz transformation), while the division of this interval into time and space parts appears to some extent
 conditional. When we change the reference frame, a part of the space projection of the interval pass into time projection
  and visa versa, approximately as in the rotating system of coordinates $OXY$ the rigid bar's $XY$-projections on a plane
can pass into one another according to certain laws. Considering the world tubes it can be said that the world tube on the
 diagram to the left is an artificial object because its vertical size is meausured in seconds, while horizontal (thickness)---
 in meters. These quantities are perfectly incomparable in Newton's classical mechanics, that's why in classical mechanics
 it is reasonable to consider the world tube, as well as the whole space of events to be a formal unity of momentous positions
 of the body or momentous spaces (a pack of sheets, separated but closely stuck together, the width and thickness of which is
  measured in different units that have absolutely no relation to each other). On the other side,  in SR the world tube is one
  four-dimensional extended object that can be measured by the interval (\ref{int}) in united units (for example, meters)
  in all directions, time as well as space. This four-dimensional body is a history of a certain three-dimensional body "frozen"\,
  in four-dimensional world. This four-dimensional body stays always extended along the time direction, thus reminding a bent thin bar.
  Every reference frame defines a combination of sections by planes of simultaneous events. The sections are three-dimensional and define
  momentous positions of a certain three-dimensional in a chosen reference frame. In another reference frame the same bar would define
  another sequence of sections. Since the appearance of the bar in four-dimensional space does not depend on the system (only the
   way the bar gets dissected in three-dimensional sections), it would be natural to call the bar {\it absolute history.}
   The sequence of the three-dimensional bodies defining {\it relative history} of a certain three-dimensional body can be
   defined only after we've chosen reference frame and set the planes of simultaneous events associated with it.
   It is absolute histories of bodies that will be the object of our study.  Note that in existing literature on SR scientist usually
 confine themselves to the consideration of one-dimensional world lines, leaving out four-dimensional extended bodies.

If we assume the four-dimensional approach and consider four-di\-men\-si\-o\-nal world bars a new physical reality of relativistic nature,
it would be natural to start by examining its four-dimensional physical qualities. Since in four-dimensional world the bars are at rest,
here we face the four-dimensional kind of statics. This statics can be reconstructed analogically to the statics of simple
three-dimensional bars.
As we know, three-dimensional bars can be subject to strains of stretching-compressing, twisting and bending \cite{land}.
In case of moderate strain these can be considered independently. Each kind of strain has its own expression for respective
elastic energy. So for the common three-dimensional bar the energy of twisting can be expressed as follows:
\begin{equation}\label{twist}
\mathcal{E}_{\text{tw}}=\int \frac{C\tau^2}{2}\, dl,
\end{equation}
where $\tau$ is the angle of the bar sections' relative turn to a unit of length (this quantity is called {\it twist}), $C$
is twist rigidity that depends on elastic constants and the form of section, and the integration is being taken along the bar axe.
In order to calculate bending energy $\mathcal{E}_{\text{b1}}$ for longitudinally unstrained bar there is a more complex expression
which we won't need. Finally, a strongly tensed bar bent by cross load has the elastic energy:
\begin{equation}\label{bend}
\mathcal{E}_{\text{b2}}=T\int dl,
\end{equation}
where $T$ is the bar's tension and the integration is taken along its axe in bent state. It should be noted that strongly tensed bars are called {\it strings.}
Unlike untensed bars, their bend resistance is defined by tension and not bend rigidity which depends on elastic constants of the bar's material. Therefore,
for strings $\mathcal{E}_{\text{b2}}\gg\mathcal{E}_{\text{b1}}.$

In order to formulate the laws of four-dimensional statics we would need to generalize the relations
mentioned above to some extent. Initially it is unclear which of the energies would be dominant for description of the four-dimensional
bars statics consistent with Newtonian mechanics as being observed in three-dimensional world. We let out technical details and present
here the result only \cite{kok}. The following conditions should be fulfilled, for equations of four-dimensional statics
in nonrelativistic limit to represent Newton's classic mechanics:
\begin{enumerate}
\item
The bars should be considered tensed strings, and the four-dimensional time-like force of tension is related to the three-dimensional mass
by relation\footnote{  It is worth reminding that in 4-dimensional world force must have the dimension of 3-dimensional energy
because simple 3-dimensional force is 4-dimensional force related to the 4-dimensional bar length unit.}: $T=mc^2.$
Thus the expression for the bend elastic energy of the tensed bar as (\ref{bend})
becomes proportional to the progressive part of action for a free mass particle $m$:
\begin{equation}\label{act1}
\mathcal{E}_{\text{b2}}=cS_{\text{progr}}=-mc^2\int ds
\end{equation}
\item
The three-dimensional  mass density $\rho$  is connected to the shear modulus $\zeta$ of the bar's four-dimensional material by a relation: $\rho c^2=\zeta.$
The four-dimensional twist energy becomes proportional to the rotational part of action for a rigid body in classical mechanics:
\begin{equation}\label{act2}
\mathcal{E}_{\text{tw}}=cS_{\text{rot}}=c\int\frac{\mathcal{J}(\omega,\omega)}{2}\,dt,
\end{equation}
where $\mathcal{J}$ is the inertia tensor, $\omega$ is angular velocity of the body rotation linked to the twisting $\tau$
of a four-dimensional bar by this relation: $\tau=\omega/c.$
\item
The proper bend rigidity of the bars is not important because the bars behave as strings.
Identification of $\zeta=\rho c^2=T/V,$ where $V$ is the three-dimensional section volume in the bar's reference frame shows
that shear rigidity (i.e. rotational inertia) is fully determined by the tension of the string.
\end{enumerate}
So, in a picture reconstructed by us {\bf the mass appears as none other than (accurate within dimension factor) time-like force that tenses
bar with such intensity that its elastic characteristics become determined by this tension.} In a four-dimensional world both simple force and mass
appear to be different projections of four-dimensional forces.

\begin{figure}[hbt]
\includegraphics{allpic.7}\caption{}\label{lapl2}
\end{figure}

In order to make clear the force nature of the mass let us turn to a known formula for Laplas pressure (see fig. \ref{lapl2}):
\begin{equation}\label{laplas}
\Delta p=2\sigma\overline{R^{-1}},
\end{equation}
that links pressure difference in gas or liquid on both sides of tensed membrane to the quantity of
local surface tension and average curvature $k=\overline{R^{-1}}=(1/R_1+1/R_2)/2$, membrane being in the state of equilibrium.
This formulae has its one-dimensional analog (see fig. \ref{lapl1})
\begin{equation}\label{laplas1}
\frac{dF}{dl}=\frac{T}{R}
\end{equation}
for normal (related to the tensed thread) bending force $F,$ force of tension of the thread $T$ and the bent thread's curvature radius $R,$
in a given point. Here $dF/dl$ stands for "one-dimensional pressure".

\begin{figure}
\includegraphics{allpic.8}\caption{}\label{lapl1}
\end{figure}
\medskip

Now it's time to show that the equation (\ref{laplas1}) is in fact somewhat simplified form of the Newton's second law.
Indeed, the four-dimensional world line velocity vector of a particle $U$ is unit, so the acceleration vector $dU/ds$
is none other than curvature vector of the world line, and its modulus equates $k=1/R$ modulus of the curvature \cite{pozdn}.
Common forces acting upon a particle are always space-like, that is, they act in a direction orthogonal $U.$
They play the role of the bending forces linear density in (\ref{laplas1}). If we rewrite the three-dimensional side of  Newton's second law in the form:
\[
\vec f=mc^2\vec k=\frac{mc^2\vec n}{R},
\]
where $\vec n$ is unit vector of the curvature direction (bar protrusion) and compare it with (\ref{laplas1}), we are now able to see that
quantity $mc^2$  indeed plays the role of the tension applied to the world line or world bar.
Note that in this picture Newton's first law becomes equivalent to a well-known statement that a strongly tensed string,
at regions where there is no bending or twisting forces, remains rectilinear. Newton's third law stays the same.
The law of mass conservation is its rough consequence for the time-like forces.

Here we are not going to discuss other interesting consequences of the four-dimensional statics which put a lot of things
in a whole new light. One can find them in the original work \cite{kok}.

\newpage

\section{The nature of Newton's third law}

In previous lecture we've already discussed the logical incompleteness of  mechanical experiments and dynamic equations
without Newton's third law. At the conclusion we've proven that the laws of three-dimensional dynamics can be rearranged
into laws of four-dimensional statics of absolute histories in four-dimensional Minkowsky space. As three-dimensional statics,
this one is also based on the laws of equilibrium (the resultant of the forces and momenta equates zero) and Newton's third law.

In order to make clear the nature of this law let us turn to axiomatics that has been developed in the works of W.Noll
and his school in the 1950-60-ies.  We'll adapt the explanation of necessary axioms of bodies and
forces for our purposes \cite{trusdell}. Axiomatics of bodies and forces contains in abstract form general characteristics of all bodies and forces
to be found in classical mechanics.
Let us consider bodies $\mathcal{A},\mathcal{B},\mathcal{C},\dots$ as  elements of some universal set $\Omega,$ called {\it the mechanical universe.}
There exist between bodies usual relations of {\it inclusion}
(for example, $\mathcal{A}\subseteq\mathcal{B}$ "body $\mathcal{A}$ is a part of body $\mathcal{B}${}"\,), of
{\it superposition} $\mathcal{A}\cap\mathcal{B}$
(common part) and {\it joint} $\mathcal{A}\cup\mathcal{B}$ ("composite body"\,), with all usual characteristics\footnote{Instead of
standard Boolean operations $\subseteq,$ $\cap,$ and $\cup$ in \cite{trusdell} was used $\preceq,$ $\wedge,$ and $\vee$ respectively, that are, in fact, more abstract.
Author give some examples, when difference between these abstract operations and Boolean ones becomes apparent. Since such examples are very exotic
from the viewpoint of application of classical mechanics, we, for the simplicity sake, will use more accustomed Boolean  operations.}.

Empty body we'll denote by the symbol $\emptyset,$ universal one by $\aleph.$ These bodies have specific characteristics:
\[
\emptyset\subseteq\mathcal{A}\quad \text{для всех}\quad \mathcal{A}\in\Omega;\quad
\mathcal{A}\subseteq\aleph \quad \text{для всех}\quad \mathcal{A}\in\Omega.
\]
If two bodies have no common part except $\emptyset,$ they are called {\it segregate.}
For any body $\mathcal{A}\in \Omega$ there is an unique body $\mathcal{A}^{\text{ext}},$ called {\it exterior} of the body $\mathcal{A},$ so that
\[
\mathcal{A}\cup\mathcal{A}^{\text{ext}}=\aleph;\quad
\mathcal{A}\cap\mathcal{A}^{\text{ext}}=\emptyset.
\]
The following relations become apparent:
\[
\emptyset^{\text{ext}}=\aleph;\quad  \aleph^{\text{ext}}=\emptyset,
\]
as well as relations
\begin{equation}\label{prop}
(\mathcal{A}^{\text{ext}})^{\text{ext}}=\mathcal{A};\quad \text{from}\quad
\mathcal{A}\subseteq\mathcal{B}\quad\text{it follows}\quad\mathcal{A}\cap\mathcal{B}^{\text{ext}}=\emptyset.
\end{equation}
The reverse for the latter is also true in the universe $\Omega$: the only bodies segregate from $\mathcal{A}^{\text{ext}}$
are the parts of the body $\mathcal{A}.$ It is easy to prove the validity of  {\it de Morgan's
relations:}
\begin{equation}\label{morgan}
(\mathcal{A}\cup\mathcal{B})^{\text{ext}}=\mathcal{A}^{\text{ext}}\cap\mathcal{B}^{\text{ext}};\quad
(\mathcal{A}\cap\mathcal{B})^{\text{ext}}=\mathcal{A}^{\text{ext}}\cup\mathcal{B}^{\text{ext}};
\end{equation}

We have an important formula of expansion:
\begin{equation}\label{decomp}
\mathcal{A}=\mathcal{B}\cup(A\cap\mathcal{B}^{\text{ext}}),
\end{equation}
for any body $\mathcal{B}\subseteq\mathcal{A}.$ And the components of the expansion are segregate:
\[
\mathcal{B}\cap(A\cap\mathcal{B}^{\text{ext}})=\mathcal{A}\cap\mathcal{B}\cap\mathcal{B}^{\text{ext}}=\emptyset.
\]

Let us now consider vector-valued functions on the pairs of segregate bodies of the  kind $\overrightarrow{F}(\mathcal{A},\mathcal{B}).$
Let us call the vector {\it a force, with which body $\mathcal{B}$ acts on body $\mathcal{A}.$}
In classical mechanics forces satisfy {\it the principles of superposition} and {\it additivity.}
Both principles are reflected in the additivity qualities of the force function, by the second and first arguments
respectively:
\begin{equation}\label{add}
\overrightarrow{F}(\mathcal{A},\mathcal{B}\cup\mathcal{C})=
\overrightarrow{F}(\mathcal{A},\mathcal{B})+\overrightarrow{F}(\mathcal{A},\mathcal{C});\quad
\overrightarrow{F}(\mathcal{B}\cup\mathcal{C},\mathcal{A})=
\overrightarrow{F}(\mathcal{B},\mathcal{A})+\overrightarrow{F}(\mathcal{C},\mathcal{A})
\end{equation}
for any bodies $\mathcal{A},\mathcal{B},\mathcal{C}$ segregated in pairs.
Assuming that in additivity relations $\mathcal{B}=\emptyset$ or $\mathcal{C}=\emptyset,$ we obtain the following relation for empty body:
\[
\overrightarrow{F}(\emptyset,\mathcal{A})=\overrightarrow{F}(\mathcal{A},\emptyset)=\overrightarrow{0}
\]
for any body  $\mathcal{A}\in\Omega.$

Let us now consider force $\overrightarrow{F}(\mathcal{A},\mathcal{A}^{\text{ext}}),$
with which the exterior   of the body $\mathcal{A}$ acts on it.
In mechanics this force is called  {\it resultant.}
Let us consider two segregate bodies $\mathcal{A}$ and $\mathcal{B}.$
The second quality (\ref{prop}) in combination with de Morgan's identities (\ref{morgan}) results in:
\begin{equation}\label{id}
\mathcal{A}^{\text{ext}}=\mathcal{B}\cup(\mathcal{A}\cup\mathcal{B})^{\text{ext}};\quad
\end{equation}
\[
\mathcal{B}^{\text{ext}}=\mathcal{A}\cup(\mathcal{A}\cup\mathcal{B})^{\text{ext}}.
\]
According to the principle of forces superposition we get:
\[
\overrightarrow{F}(\mathcal{A},\mathcal{A}^{\text{ext}})=
\overrightarrow{F}(\mathcal{A},\mathcal{B})+
\overrightarrow{F}(\mathcal{A},(\mathcal{A}\cup\mathcal{B})^{\text{ext}});\quad
\]
\[
\overrightarrow{F}(\mathcal{B},\mathcal{B}^{\text{ext}})=
\overrightarrow{F}(\mathcal{B},\mathcal{A})+
\overrightarrow{F}(\mathcal{B},(\mathcal{A}\cup\mathcal{B})^{\text{ext}}).
\]
Adding the equations together with the use of the forces additivity principle,
and after we've gathered the expressions of exterior on the right side, the following equation can be obtained:
\begin{equation}\label{forces}
\overrightarrow{F}(\mathcal{A},\mathcal{B})+\overrightarrow{F}(\mathcal{B},\mathcal{A})=
\overrightarrow{F}(\mathcal{A},\mathcal{A}^{\text{ext}})+
\overrightarrow{F}(\mathcal{B},\mathcal{B}^{\text{ext}})-
\overrightarrow{F}(\mathcal{A}\cup\mathcal{B},(\mathcal{A}\cup\mathcal{B})^{\text{ext}})
\end{equation}
It is the main identity necessary to analyze the nature of Newton's third law.
From relation (\ref{forces}) it follows that in the mechanical universe, where the force principles of
superposition and additivity are fulfilled, {\bf Newton's third law takes place only when the resultant also is an additive
function of the first argument by segregate bodies}:
\begin{equation}\label{noll}
\overrightarrow{F}(\mathcal{A},\mathcal{B})+\overrightarrow{F}(\mathcal{B},\mathcal{A})=\overrightarrow{0}\Leftrightarrow
\overrightarrow{F}(\mathcal{A}\cup\mathcal{B},(\mathcal{A}\cup\mathcal{B})^{\text{ext}})=
\overrightarrow{F}(\mathcal{A},\mathcal{A}^{\text{ext}})+
\overrightarrow{F}(\mathcal{B},\mathcal{B}^{\text{ext}})
\end{equation}
for all $\mathcal{A}\in\Omega,$ $\mathcal{B}\in\Omega$ and
$\mathcal{A}\cap\mathcal{B}=\emptyset.$
This statement is the essence of {\it Noll's theorem.} Let us define the expression $\overrightarrow{F}(\mathcal{A},\mathcal{B})+\overrightarrow{F}(\mathcal{B},\overrightarrow{A})$
through $\overrightarrow{\Delta}(\mathcal{A},\mathcal{B})$ and call it {\it discrepancy of forces for bodies $\mathcal{A}$ and $\mathcal{B}$.}
Noll's theorem states that discrepancy is the measure of nonadditivity concerning interactions of the composite body with its environment.
Then let us turn back to statics, where for any $\mathcal{A}$ we have $\overrightarrow{F}(\mathcal{A},\mathcal{A}^{\text{ext}})=0.$
It follows from the Noll's theorem that in
the world of statics the discrepancy of forces for any pair of bodies is identical  zero.
In other words, {\it in statics Newton's third law is in effect according to the general principles of forces superposition and additivity.}
It is easy to notice that Newton's third law is stronger than any principle of superposition or additivity taken independently.
Indeed, if we apply the third law to any of the items under condition (\ref{add}),
we can see that the condition of additivity becomes principle of superposition and visa versa, by any three bodies segregated in pairs.
That means the validity of the third law and one of the principles result in the validity for the second principle while the third law
follows from the principles of superposition and additivity only on an extra condition that is additivity of the resultant.
This condition in its turn does not follow from the principles of superposition and additivity.

Let us now try to generalize the formulations in order to consider the forces of interactions and principles of superposition and
additivity not only on segregate bodies. First we'll redefine relations (\ref{add}) on bodies with the superposition that differs from zero body:
\begin{equation}\label{add1}
\overrightarrow{F}(\mathcal{A},\mathcal{B}\cup\mathcal{C})=
\overrightarrow{F}(\mathcal{A},\mathcal{B})+\overrightarrow{F}(\mathcal{A},\mathcal{C})-
\overrightarrow{F}(\mathcal{A},\mathcal{B}\cap\mathcal{C});
\end{equation}
\begin{equation}\label{add2}
\overrightarrow{F}(\mathcal{B}\cup\mathcal{C},\mathcal{A})=
\overrightarrow{F}(\mathcal{B},\mathcal{A})+\overrightarrow{F}(\mathcal{C},\mathcal{A})-
\overrightarrow{F}(\mathcal{B}\cap\mathcal{C},\mathcal{A}).
\end{equation}

Under $\mathcal{B}\cap\mathcal{C}=\emptyset$ these relations transform into (\ref{add})
and in fact take into consideration that the force interaction $\mathcal{B}\cap\mathcal{C}$ if it differs from zero,
has been twice accounted for in the formulae.

Let us now consider the force of interaction on bodies $\mathcal{A}$ and $\mathcal{B}$ with $\mathcal{A}\cap\mathcal{B}=\mathcal{C}\neq\emptyset.$
Using the representations:
\begin{equation}\label{dec}
\mathcal{A}=\mathcal{A}'\cup(\mathcal{A}\cap\mathcal{B});\quad
\mathcal{B}=\mathcal{B}'\cup(\mathcal{A}\cap\mathcal{B});
\end{equation}
where $\mathcal{A}'=\mathcal{A}\setminus\mathcal{B}$ and $\mathcal{B}'=\mathcal{B}\setminus\mathcal{A}$ are
{\it bumps} of $\mathcal{A}$ above $\mathcal{B}$ and $\mathcal{B}$ above $\mathcal{A}$ respectively,
(all the characteristics of theoretical set $\setminus$-operation taken into account), we  define the interaction force of non-segregate bodies this way:
\[
\overrightarrow{F}(\mathcal{A},\mathcal{B})=
\overrightarrow{F}(\mathcal{A}'\cup(\mathcal{A}\cap\mathcal{B}),
\mathcal{B}'\cup(\mathcal{A}\cap\mathcal{B}))=
\]
\begin{equation}\label{forceg}
\overrightarrow{F}(\mathcal{A}',\mathcal{B}')+\overrightarrow{F}(\mathcal{A}',\mathcal{A}\cap\mathcal{B})+
\overrightarrow{F}(\mathcal{A}\cap\mathcal{B},\mathcal{B}')+\overrightarrow{F}(\mathcal{A}\cap\mathcal{B},\mathcal{A}\cap\mathcal{B}).
\end{equation}
In the case of segregate bodies our definition transforms into identity of the kind $\overrightarrow{F}(\mathcal{A},\mathcal{B})=
\overrightarrow{F}(\mathcal{A},\mathcal{B}).$
In the case of non-segregate it reads thus: {\it the force with which body $\mathcal{B}$ acts on the not segregated from it
body $\mathcal{A}$ is formed by the force with which the bump $\mathcal{B}'$ acts on the bump $\mathcal{A}'$
plus the force of the superposition on bump $\mathcal{A}'$, force of $\mathcal{B}'$ on superposition and
force of superposition self-action.}
The three first components are simple forces acting on segregated bodies, meaning that all that is new in the forces
interactions with non-segregate bodies lies in the quantities of the self-acting forces of the kind:
$\overrightarrow{F}(\mathcal{A},\mathcal{A})\equiv\overrightarrow{\mathfrak{F}}(\mathcal{A}).$

Let us make clear the nature of the force. First we'll examine the interaction force of a certain body
$\mathcal{A}$ with the universal body $\aleph.$ In accordance to our definition (\ref{forceg}) we get:
\[
\overrightarrow{F}(\mathcal{A},\aleph)=\overrightarrow{F}(\mathcal{A},\mathcal{A}^{\text{ext}})+
\overrightarrow{\mathfrak{F}}(\mathcal{A}).
\]
Therefore, the self-acting force can be described as difference:
\[
\overrightarrow{\mathfrak{F}}(\mathcal{A})=
\overrightarrow{F}(\mathcal{A},\aleph)-\overrightarrow{F}(\mathcal{A},\mathcal{A}^{\text{ext}})
\]
of the force of interaction with universal body and the resultant force.
Since all the common bodies in classical mechanics are in a sense "small bodies"\, when compared to the environment
as well as the universal body, we have\footnote{In order to compare bodies on $\Omega$
we need to introduce a measure. For simple geometrical measure --- volume and physical measure ---
mass the conditions of smallness are reduced to strong inequalities $V(\mathcal{A})\ll V(\mathcal{A}^{\text{ext}})$ or
$M(\mathcal{A})\ll M(\mathcal{A}^{\text{ext}}),$ which are almost always valid.} $\mathcal{A}^{\text{ext}}\approx\aleph$ and therefore
\[
\overrightarrow{\mathfrak{F}}(\mathcal{A})\approx0.
\]
However, the self-acting force can appear resulting from incomplete compensation  of two approximately equal values.

Now let us show that the mechanical universe with additive resultant and non-trivial interaction of non-segregate bodies is impossible.
In order to do that we'll need to generalize identities (\ref{id}),
considering the case of non-segregate bodies and put to use the principle of superposition for the bumps.
It is easy to prove that for non-segregate bodies $\mathcal{A}$ and $\mathcal{B}$ identities (\ref{id})
acquire the form:
\begin{equation}\label{id1}
\mathcal{A}^{\text{ext}}=\mathcal{B}\cup(\mathcal{A}\cup\mathcal{B})^{\text{ext}}\setminus(\mathcal{A}\cap\mathcal{B});\quad
\mathcal{B}^{\text{ext}}=\mathcal{A}\cup(\mathcal{A}\cup\mathcal{B})^{\text{ext}}\setminus(\mathcal{A}\cap\mathcal{B}).
\end{equation}
The principle of superposition for the bumps, in accordance to their definition, has the form:
\begin{equation}\label{add3}
\overrightarrow{F}(\mathcal{A},\mathcal{B}\setminus\mathcal{C})=
\overrightarrow{F}(\mathcal{A},\mathcal{B}\cup\mathcal{C})
-\overrightarrow{F}(\mathcal{A},\mathcal{C}).
\end{equation}
Then, after we've gone through the calculations analogous to those made under the conclusion (\ref{forceg})
and with regard to (\ref{id1})
and (\ref{add3}), we obtain:
\begin{equation}\label{forcesg}
\overrightarrow{F}(\mathcal{A},\mathcal{B})+\overrightarrow{F}(\mathcal{B},\mathcal{A})-
\overrightarrow{F}(\mathcal{A}\cup\mathcal{B},\mathcal{A}\cap\mathcal{B})-
\overrightarrow{\mathfrak{F}}(\mathcal{A}\cap\mathcal{B})=
\end{equation}
\[\overrightarrow{F}(\mathcal{A},\mathcal{A}^{\text{ext}})+
\overrightarrow{F}(\mathcal{B},\mathcal{B}^{\text{ext}})-
\overrightarrow{F}(\mathcal{A}\cup\mathcal{B},(\mathcal{A}\cup\mathcal{B})^{\text{ext}})-
\overrightarrow{F}(\mathcal{A}\cap\mathcal{B},(\mathcal{A}\cup\mathcal{B})^{\text{ext}})
\]
the relation which generalizes (\ref{forces}) for the case $\mathcal{A}\cap\mathcal{B}\neq\emptyset.$
From it directly follows Noll's generalized theorem: {\it  additivity of the resultant on all bodies is equivalent to the expression for
discrepancy:}
\begin{equation}\label{d}
\overrightarrow{\Delta}(\mathcal{A},\mathcal{B})=\overrightarrow{F}(\mathcal{A}\cup\mathcal{B},\mathcal{A}\cap\mathcal{B})+
\overrightarrow{\mathfrak{F}}(\mathcal{A}\cap\mathcal{B}).
\end{equation}
Assuming that the resultant is additive on all bodies, let us now examine the expression obtained in detail. With the help of the decomposition
(\ref{dec}) and upon denoting $\mathcal{A}\cap\mathcal{B}=\mathcal{C},$ after some elementary redenotations  we obtain from (\ref{d}):
\begin{equation}\label{eq}
\overrightarrow{\Delta}(\mathcal{A}',\mathcal{B}')+
\overrightarrow{F}(\mathcal{C},\mathcal{A}'\cup\mathcal{B}')=\overrightarrow{0}.
\end{equation}
Bodies $\mathcal{A}',\mathcal{B}'$ and $\mathcal{C}$ can be considered arbitrary segregate bodies, so for  $\mathcal{A}'=\emptyset$ (\ref{eq}) gives us:
\[
\overrightarrow{F}(\mathcal{C},\mathcal{B}')=\overrightarrow{0}.
\]
for all segregate bodies. Therefore, {\it in a universe where there is interaction between  non-segregate bodies and the
resultant is additive, segregate bodies do not interact at all (hence the resultant equates zero),
while non-segregate interact by means of self-acting force of its common part.}
For such system of forces $\overrightarrow{F}(\mathcal{A},\mathcal{B})=\overrightarrow{F}(\mathcal{B},\mathcal{A})=
\overrightarrow{\mathfrak{F}}(\mathcal{A}\cap\mathcal{B}).$ Besides, self-acting force has additive quality:
\[
\overrightarrow{\mathfrak{F}}(\mathcal{A}_1\cup\mathcal{A}_2)
=\overrightarrow{\mathfrak{F}}(\mathcal{A}_1)+\overrightarrow{\mathfrak{F}}(\mathcal{A}_2)
\]
for any subdivision $\mathcal{A}=\mathcal{A}_1\cup\mathcal{A}_2$
с $\mathcal{A}_1\cap\mathcal{A}_2=\emptyset.$

In fact, in a world of self-acting the role of resultant would be played by full force $\overrightarrow{F}(\mathcal{A},\aleph),$
not by the resultant $\overrightarrow{F}(\mathcal{A},\mathcal{A}^{\text{ext}}).$
The condition of this full force additivity will have the form:
\[
\overrightarrow{F}(\mathcal{A},\aleph)+
\overrightarrow{F}(\mathcal{B},\aleph)-
\overrightarrow{F}(\mathcal{A}\cup\mathcal{B},\aleph) - \overrightarrow{F}(\mathcal{A}\cap\mathcal{B},\aleph)=\overrightarrow{0}.
\]
Adding to and subtracting from the right side (\ref{forcesg})
necessary components of self-acting, of the kind $\overrightarrow{F}(\mathcal{A},\mathcal{A}),\dots,$
and after all necessary simplifyings we would come to the same conclusion: {\it in a universe where the full force is additive there is no
interaction between segregate bodies while non-segregate interact only as a result of their supposition self-acting.}
It is apparent that in such a world consistent  statics would be impossible, because if $\overrightarrow{F}(\mathcal{A},\aleph)=\overrightarrow{0}$
for any $\mathcal{A},$ then $\mathfrak{F}(\mathcal{A})=\overrightarrow{0}$ for any $\mathcal{A},$ meaning forces vanish at all.

\begin{figure}[hbt]
\includegraphics{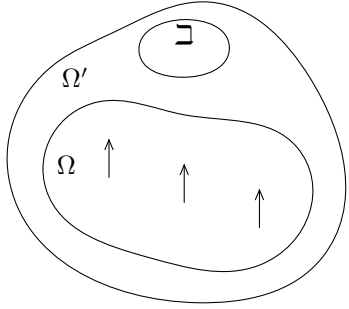}\caption{\small Universe $\Omega$ with self-acting looks similar to universe $\Omega'=\Omega\cup\{\beth\}$
without self-acting. Instead of the acceleration force
there is the force of interaction between bodies $\Omega$ and body $\beth,$ which is inaccessible for observation. For that reason this interaction will
be viewed as additive self-acting  by observers from $\Omega$
}\label{univ}
\end{figure}

It is interesting that the unusual universe we got can be interpreted differently. Let us consider a broader universe $\Omega'=\Omega\cup\{\beth\},$
where bodies from  $\Omega$ do not interact between themselves but do interact with body
$\beth$ (Fig.\ref{univ}).
Assuming $\overrightarrow{\mathfrak{F}}(\mathcal{A})\equiv\overrightarrow{F}(\mathcal{A},\beth)$ for all $\mathcal{A}\in\Omega,$ where $\overrightarrow{F}$
satisfies only the principle of additivity.
We can see that in such extended universe the observer, being the part of $\Omega,$ perceive the force of interaction
with body $\beth,$ that lies outside the universe $\Omega$ as self-acting.
While experimenting with body $\beth$ in universe $\Omega$ are fundamentally impossible,
the investigator can always choose between the terms of "self-acting" or "transcendental" interacting, according to personal philosophical beliefs!

\bigskip
I am grateful to Evgeniy Stern for assistance with figures  and to
Anna Turuntaeva for translation of the text in English.

\end{document}